\documentclass[aps,pre,twocolumn,superscriptaddress,nofootinbib]{revtex4-2}
\usepackage[T1]{fontenc}
\usepackage[utf8]{inputenc}
\usepackage{amsmath,amssymb}
\usepackage{graphicx}
\usepackage[caption=false]{subfig}
\usepackage{booktabs}
\usepackage{xcolor}
\usepackage{hyperref}
\usepackage{tikz}

\newif\iffinal
\iffinal
  
  \newcommand{\note}[1]{}
\else
  
  \newcommand{\note}[1]{\par\smallskip\noindent\fcolorbox{red}{red!5}{%
    \parbox{\dimexpr\linewidth-2\fboxsep-2\fboxrule}{\small\textbf{Note:} #1}}\par\smallskip}
\fi
\graphicspath{{figures/}}
\begin{document}

\title{Basin Adjacency Graphs: A New Approach to Characterizing Basins of Attraction} 
\author{Alexandre Wagemakers} \affiliation{Nonlinear Dynamics, Chaos and Complex Systems Group, Departamento de F\'isica, Universidad Rey Juan Carlos, Tulip\'an s/n, 28933 M\'ostoles, Madrid, Spain} \author{Aryan Srivastava} 
\affiliation{School of Mathematical Sciences, UM-DAE Centre for Excellence in Basic Sciences (CEBS), University of Mumbai, Vidyanagari, Mumbai 400098, India} 
\author{Miguel A.~F. Sanju\'an} \affiliation{Nonlinear Dynamics, Chaos and Complex Systems Group, Departamento de F\'isica, Universidad Rey Juan Carlos, Tulip\'an s/n, 28933 M\'ostoles, Madrid, Spain} \affiliation{Royal Academy of Sciences of Spain, Valverde 22, 28004 Madrid, Spain}

\begin{abstract}
Fractal basins of attraction are commonly characterized through resolution-dependent measurements, such as box-counting methods, uncertainty exponents, and basin entropy. Here we introduce a different approach based on the adjacency relations between connected components of the basins, without retaining their detailed spatial geometry. A finite-resolution basin diagram is represented by a graph with one vertex per connected component and edges linking adjacent components belonging to different basins. We investigate how the number of vertices scales with the spatial resolution $N$, following a power law $N^\alpha$ in the systems considered. Under suitable assumptions, the exponent $\alpha$ is related to the Kor\v{c}\'ak exponent of the component-size distribution and can coincide with the box-counting dimension of the basin boundary when its fractal structure arises from the proliferation of components. The graph also provides adjacency-based diagnostics that may help characterize basin configurations exhibiting the Wada property. We illustrate the approach using idealized Sierpi\'nski constructions, the Newton and relaxed Newton families, and three dynamical systems with previously established Wada basins.
\end{abstract}

\maketitle

\begin{quotation}
\noindent\textbf{%
A nonlinear system can approach different final states depending on its initial conditions. The regions leading to these states, called basins of attraction, often form intricate patterns that make prediction difficult. We describe these patterns using a graph in which each connected basin region becomes a vertex and neighboring regions are linked by edges. This representation reveals how basin regions are organized and how additional components emerge as the spatial resolution increases. For several benchmark systems, the growth of the number of vertices gives an exponent close to independently estimated fractal boundary dimensions. Comparisons with the distribution of component areas show when these descriptions agree and how they differ for thin or highly fragmented regions. The graph also records which basins neighbor each component, providing a complementary view of the spatial organization underlying sensitivity to initial conditions.%
}
\end{quotation}

\section{Introduction}

Characterizing the basins of attraction of a multistable dynamical system provides valuable information about its long-term behavior and its sensitivity to initial conditions~\cite{daza2024pt}. Several methods have been developed to quantify the geometry and complexity of basin boundaries, including the uncertainty exponent, the box-counting dimension, and basin entropy and its variants~\cite{grebogi1983,mcdonald1985,aguirre2009,daza2016,daza2022,gusso2021}. These approaches rely on the analysis of initial conditions at different spatial resolutions, often using boxes of characteristic size $\varepsilon$ to investigate properties such as the coexistence of different final states within a small region of phase space. Although each method has its own limitations, they have been successfully applied to a wide variety of nonlinear dynamical systems.

In this article, we propose a different approach based on the adjacency relations between connected components of the basins of attraction. Rather than describing their detailed spatial geometry, we represent a finite-resolution basin diagram by a graph. The construction is inspired by region-adjacency graphs used in image analysis~\cite{tremeau2000}, in which each region is represented by a vertex and neighboring regions are connected by edges. Applied to two-dimensional basins of attraction, this leads to what we call the \emph{basin adjacency graph} (BAG). Graph-based approaches are not new in nonlinear dynamics. Recurrence networks and related constructions encode time series as networks whose properties reflect the geometry of attractors~\cite{donner2010,zou2019}, while the qualitative behavior of a map can be represented through a graph of its chain-recurrent classes~\cite{deleo2021}. The BAG differs from these approaches in that it describes the adjacency and connectivity of basin components rather than the structure of the attractors themselves.

When the fractal structure of basin boundaries is accompanied by a proliferation of disconnected components, increasing the spatial resolution reveals an increasingly complex graph, while large connected regions are represented by single vertices. This reduction allows us to investigate how the number of graph vertices changes with resolution. In particular, we study the scaling of the number of vertices and its relationship with the Kor\v{c}\'ak exponent of the component-size distribution, originally introduced in the study of island sizes~\cite{mandelbrot1982} and subsequently used to characterize the spatial patchiness of landscapes~\cite{hastings1982,imre2012}. Under suitable assumptions on the size distribution and geometry of the components, the vertex-scaling exponent is related to the Kor\v{c}\'ak exponent and may coincide with the box-counting dimension of the basin boundary when its fractal complexity arises from the proliferation of components. We examine this relationship using idealized Sierpi\'nski constructions and several dynamical systems, including Newton's method. In addition, we investigate whether the adjacency structure provides useful information about basin configurations exhibiting the Wada property~\cite{kennedy1991,nusse1996,daza2015,wagemakers2021}.

An important limitation of this approach is that not every fractal boundary is associated with a proliferation of connected components. For example, the Julia sets of $z^2+c$, with $c$ in the main cardioid of the Mandelbrot set, may form fractal boundaries separating two connected regions. In such cases, the adjacency graph contains only two vertices, independently of the fractal dimension of the boundary. Similar limitations arise for geometrically irregular boundaries separating connected regions, such as coastlines. The graph therefore does not capture every aspect of boundary geometry, and its scaling properties must be interpreted accordingly. A related distinction arises for the Wada property. The classical Lakes of Wada construction originates in topology~\cite{yoneyama1917}, whereas its importance for basins of attraction in dynamical systems was established by Kennedy and Yorke~\cite{kennedy1991}. The adjacency graph alone cannot distinguish a Wada configuration from three mutually adjacent regions with ordinary, non-Wada boundaries. Consequently, graph adjacency provides useful structural information but does not, by itself, establish the Wada property. We focus here on systems in which the proliferation and adjacency of basin components provide additional information about their geometrical and dynamical organization.

\section{The basin adjacency graph}

We first consider a basin diagram obtained by sampling $N\times N$ initial conditions on a regular two-dimensional grid. Each point is labeled according to the attractor, among the $N_A$ attractors of the system, to which its trajectory converges. The construction begins with a pixel graph in which each point is connected to its eight nearest neighbors, including horizontal, vertical, and diagonal neighbors, except at the edges of the grid. By contracting all edges connecting points belonging to the same basin, each connected component is reduced to a single vertex, while edges between components of different basins are retained.

Formally, let $C_1,\dots,C_M$ denote the connected components obtained by retaining only those edges of the pixel graph that join points with the same basin label. The basin adjacency graph (BAG) is defined as $G=(V,E)$, where $V=\{C_1,\dots,C_M\}$ and $(C_i,C_j)\in E$ if and only if $C_i$ and $C_j$ belong to different basins and at least one pixel of $C_i$ is an eight-neighbor of a pixel of $C_j$. Each vertex retains the label of its corresponding basin and is assigned a weight $A_i$, representing the number of pixels in the component. The resulting graph is simple and undirected, and its construction requires $O(N^2)$ operations, excluding the computation and classification of the trajectories.

The choice of pixel connectivity deserves some attention. In particular, eight-neighbor connectivity may join regions that touch only diagonally, and the resulting graph is not necessarily planar. Different connectivity conventions may therefore affect the number of components and their adjacency relations. These effects should be taken into account when interpreting the graph, especially at finite resolution~\cite{rosenfeld1979}.

From $G$, we extract the number of vertices $|V|$, the degree $k_i$ of each vertex, and the \emph{basin-neighbor count} $w(C_i)$, defined as the number of distinct basin labels, other than its own, represented among the neighbors of $C_i$. The value $w(C_i)=N_A-1$ means that $C_i$ is adjacent to at least one component of every other basin. This quantity provides information about the adjacency structure of the basins, but it does not by itself establish the Wada property, which requires every boundary point of a basin to belong to the boundaries of all the other basins. The distinction is important because different basins may be mutually adjacent without sharing the same boundary. The basin-neighbor count should therefore be regarded as an adjacency-based descriptor rather than a sufficient test for the Wada property~\cite{daza2015,wagemakers2021}.

Although we focus here on regular grids, the construction could in principle be extended to nonuniform samplings by replacing the pixel graph with a suitable neighborhood graph, such as a $k$-nearest-neighbor graph. Such an extension would require examining the dependence of the resulting connectivity and adjacency relations on the chosen neighborhood definition.

\section{Vertex scaling and the Korčák dimension}

Our goal is to extract from the BAG an exponent that characterizes the proliferation of resolved basin components as the spatial resolution increases. To this end, we construct the graph at different resolutions $\varepsilon$ and examine the dependence of its number of vertices on $\varepsilon$. When a power law is observed, we define the vertex-scaling exponent $\alpha$ through $|V(\varepsilon)|\sim\varepsilon^{-\alpha}$. For a finite partition whose connected regions are adequately resolved, the vertex count becomes constant, giving $\alpha=0$. When finer sampling reveals additional disconnected components, the vertex count may increase. Such growth is not a general consequence of boundary fractality, since a fractal boundary can separate a finite number of connected regions, e.g., the conected Julia set.

The interpretation of $\alpha$ can be related to the distribution of component areas. When this distribution follows a Kor\v{c}\'ak law, the number of components with area greater than $a$ satisfies~\cite{mandelbrot1982,hastings1982,imre2012}
\begin{equation}
  \mathcal{N}(A>a) \propto a^{-K}.
  \label{eq:korcak}
\end{equation}
For sufficiently compact components, whose characteristic linear size scales as the square root of their area, the area threshold for resolution is of order $\varepsilon^2$. Assuming that components above this threshold are resolved separately and that the power law holds over the relevant range of scales, we obtain
\begin{equation}
  |V(\varepsilon)| \simeq \mathcal{N}(A>\varepsilon^2) \propto \varepsilon^{-2K},
  \qquad \alpha = 2K \equiv D_K ,
  \label{eq:alpha_korcak}
\end{equation}
where $D_K$ denotes the exponent associated with the Kor\v{c}\'ak area distribution. It should not generally be identified with the dimension of the basin boundary. The box-counting and Hausdorff dimensions characterize the geometry of a set~\cite{farmer1983,theiler1990}, whereas $\alpha$ characterizes the proliferation of resolved components. A boundary separating a finite number of connected regions may be fractal while the corresponding idealized graph has $\alpha=0$. The exponents can coincide in self-similar constructions where the hierarchy of components and the boundary geometry have the same scaling~\cite{mandelbrot1982,imre2012}. For thin or elongated components, however, resolution depends on their width and separation as well as their area, so Eq.~\eqref{eq:alpha_korcak} may fail.

As an idealized illustration, we consider the Sierpi\'nski triangle at iteration level $m$. For this construction, we treat filled subtriangles that meet only at isolated points as separate regions. Each filled subtriangle and each removed triangular hole is represented by a vertex. Edges join filled subtriangles to holes when they share a boundary segment of positive length. The exterior of the enclosing triangle is excluded.

This convention differs from ordinary topological connectivity, under which the filled set is connected through its retained corner points. It also differs from the eight-neighbor pixel connectivity used in the numerical BAG construction.

Let $k_m$ denote the number of filled subtriangles and $b_m$ the number of holes. The construction gives
\begin{equation}
    k_m = 3^m, \qquad b_m = \sum_{j=0}^{m-1} 3^j = \frac{3^m - 1}{2},
\end{equation}
The number of graph vertices is
\begin{equation}
    |V_m| = k_m + b_m = \frac{3^{m+1} - 1}{2}.
\end{equation}
For a triangle with unit side, the smallest geometrical scale at level $m$ is $\varepsilon=2^{-m}$. Expressing the vertex count in terms of this scale, we obtain for large $m$
\begin{equation}
    |V(\varepsilon)| \simeq \frac{3}{2}\, 3^m = \frac{3}{2}\, \varepsilon^{-\log 3/\log 2} = \frac{3}{2}\, \varepsilon^{-d_f},
\end{equation}
where $d_f=\log 3/\log 2\approx1.585$ is the Hausdorff dimension of the limiting Sierpi\'nski triangle. The equality $\alpha=d_f$ follows because the vertex count grows asymptotically as $3^m$, while the characteristic length decreases as $2^{-m}$. If point contacts were instead retained as connections, the entire filled set would contribute one vertex, giving $|V_m|=(3^m+1)/2$. This alternative count changes the asymptotic prefactor but preserves the exponent $\alpha=d_f$.

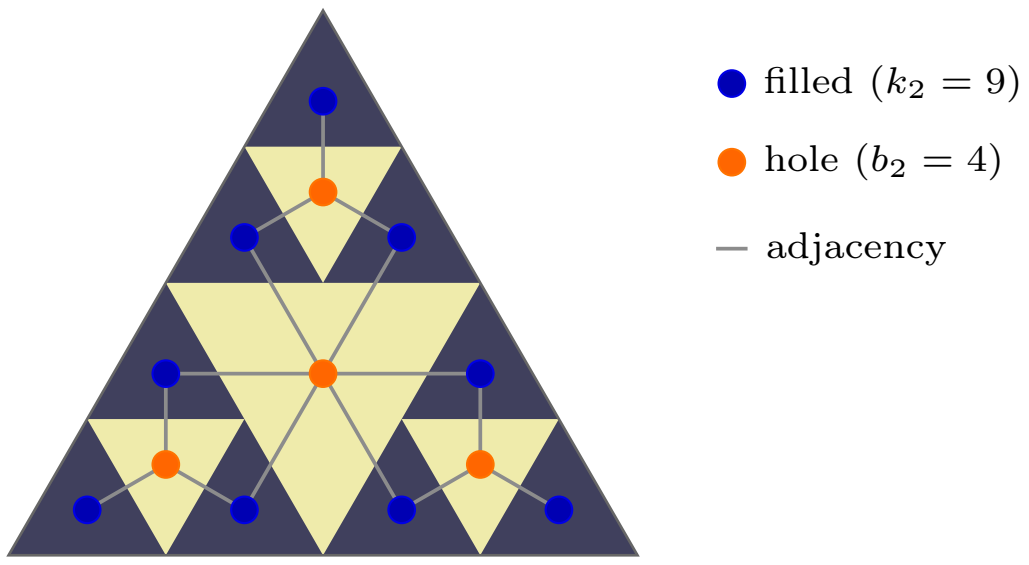
\begin{figure}[ht]
\centering
\begin{tikzpicture}[
    bnode/.style={circle, fill=blue!70!black, draw=blue!90!black, line width=0.4pt, inner sep=0pt, minimum size=5pt},
    wnode/.style={circle, fill=orange!80!red, draw=orange!90!red, line width=0.4pt, inner sep=0pt, minimum size=5pt},
    gedge/.style={draw=black!45, line width=0.7pt},
    scale=0.52
]
\fill[blue!15!black!85, opacity=0.88] (0,0) -- (8,0) -- (4,6.93) -- cycle;
\fill[orange!10!yellow!30, opacity=0.92] (4,0) -- (2,3.465) -- (6,3.465) -- cycle;
\fill[orange!10!yellow!30, opacity=0.92] (2,0) -- (1,1.7325) -- (3,1.7325) -- cycle;
\fill[orange!10!yellow!30, opacity=0.92] (6,0) -- (5,1.7325) -- (7,1.7325) -- cycle;
\fill[orange!10!yellow!30, opacity=0.92] (4,3.465) -- (3,5.1975) -- (5,5.1975) -- cycle;
\draw[black!60, line width=0.5pt] (0,0) -- (8,0) -- (4,6.93) -- cycle;
\coordinate (b1) at (1,    0.5775);
\coordinate (b2) at (3,    0.5775);
\coordinate (b3) at (2,    2.31);
\coordinate (b4) at (5,    0.5775);
\coordinate (b5) at (7,    0.5775);
\coordinate (b6) at (6,    2.31);
\coordinate (b7) at (3,    4.0425);
\coordinate (b8) at (5,    4.0425);
\coordinate (b9) at (4,    5.775);
\coordinate (w1) at (4,    2.31);
\coordinate (w2) at (2,    1.155);
\coordinate (w3) at (6,    1.155);
\coordinate (w4) at (4,    4.62);
\draw[gedge] (w2)--(b1); \draw[gedge] (w2)--(b2); \draw[gedge] (w2)--(b3);
\draw[gedge] (w3)--(b4); \draw[gedge] (w3)--(b5); \draw[gedge] (w3)--(b6);
\draw[gedge] (w4)--(b7); \draw[gedge] (w4)--(b8); \draw[gedge] (w4)--(b9);
\draw[gedge] (w1)--(b2); \draw[gedge] (w1)--(b3); \draw[gedge] (w1)--(b4);
\draw[gedge] (w1)--(b6); \draw[gedge] (w1)--(b7); \draw[gedge] (w1)--(b8);
\foreach \i in {1,...,9} { \node[bnode] at (b\i) {}; }
\foreach \i in {1,...,4} { \node[wnode] at (w\i) {}; }
\node[bnode, label={[font=\scriptsize]right:filled ($k_2 = 9$)}] at (9.2, 6.0) {};
\node[wnode, label={[font=\scriptsize]right:hole ($b_2 = 4$)}]    at (9.2, 5.0) {};
\draw[gedge] (9.0, 3.9) -- (9.4, 3.9) node[right, font=\scriptsize]{adjacency};
\end{tikzpicture}
\caption{Idealized region-adjacency graph of the
Sierpi\'nski triangle at level $m=2$.
Each of the nine filled subtriangles (blue) and
four triangular holes (orange) is represented by
a vertex. Filled subtriangles meeting only at
isolated points are treated as separate regions.
Edges represent shared boundary segments of positive
length between filled subtriangles and holes.
The graph contains $|V|=13$ vertices and $|E|=15$ edges.
This idealized convention differs from the
eight-neighbor pixel connectivity used for the
numerical BAGs.}
\label{fig:sierpinski}
\end{figure}

The Sierpi\'nski triangle thus illustrates how component proliferation can yield an exponent equal to a geometrical fractal dimension (Fig.~\ref{fig:sierpinski}). For the Sierpi\'nski carpet, counting its connected filled region as one vertex and each removed square as a separate vertex gives $|V_m|=1+(8^m-1)/7$. With characteristic scale $\varepsilon=3^{-m}$, this yields $\alpha=\log 8/\log 3\approx1.893$, again equal to the Hausdorff dimension of the limiting fractal.

More generally, a hierarchy generating $a^j$ additional components of characteristic linear size $b^{-j}$ at level $j$, with $a>1$ and $b>1$, has a cumulative component count growing asymptotically as $a^m$. If these components are resolved at scale $\varepsilon\sim b^{-m}$, the resulting exponent is $\alpha=\log a/\log b$. Identifying this exponent with a boundary dimension requires additional assumptions about the geometry of the hierarchy. The relation $\alpha=2K$ also requires the area-to-length scaling assumed in Eq.~\eqref{eq:alpha_korcak}.

\section{Results}

\subsection{Newton's method}

\begin{figure}[t]
    \centering
    \subfloat[Basins, $N=200$]{
        \includegraphics[width=0.42\linewidth]{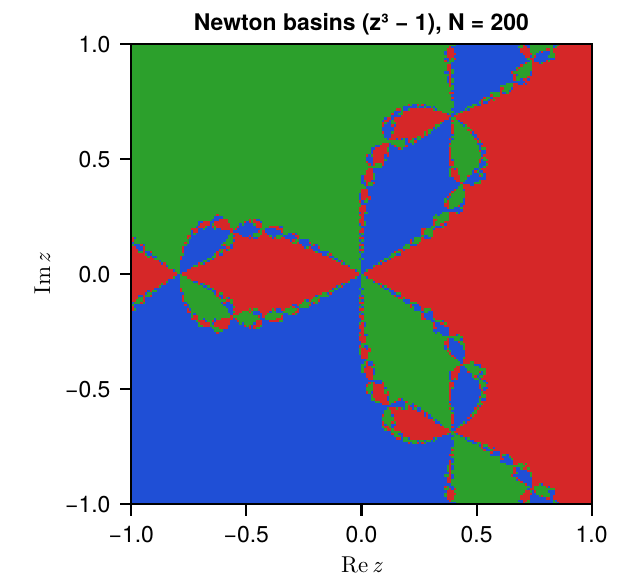}
    }\hfill
    \subfloat[Vertex scaling]{
        \includegraphics[width=0.48\linewidth]{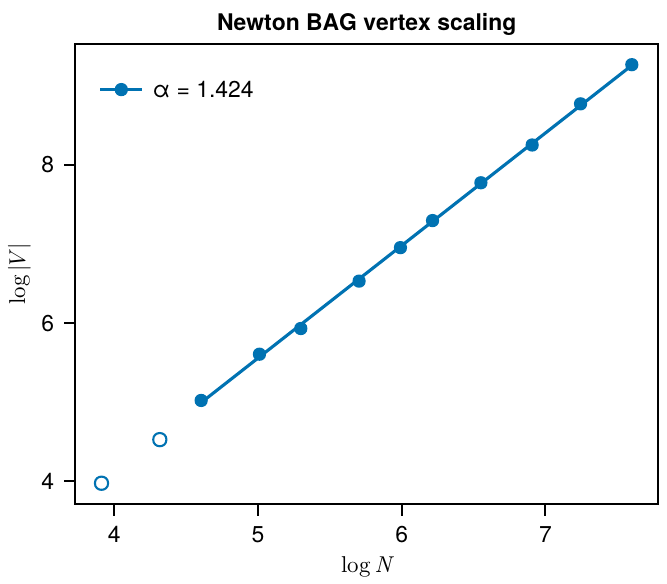}
    }
    \caption{(a) Basins of attraction of Newton's method applied to $z^3-1$.
    (b) Scaling of the number of vertices of the basin adjacency graph with the grid resolution $N$. A log-log fit over $100\leq N\leq2000$ gives $\alpha=1.424$, consistent with independent estimates of the fractal boundary dimension. Open symbols ($N<100$) are excluded from the fit.}
    \label{fig:newton}
\end{figure}

Newton's method applied to $z^3-1$,
\[
z\mapsto z-\frac{z^3-1}{3z^2},
\]
has three attracting fixed points corresponding to the cube roots of unity. Their basins of attraction provide a well-known example of the Wada property~\cite{daza2015}. Figure~\ref{fig:newton}(a) shows the basins on the square $[-1,1]\times[-1,1]$, corresponding to the computational domain considered by Epureanu and Greenside~\cite{epureanu1998}.

Since the pixel size scales as $\varepsilon\sim N^{-1}$, the vertex-scaling exponent $\alpha$ can be estimated from the slope of $\log|V|$ versus $\log N$. Figure~\ref{fig:newton}(b) illustrates this dependence. A linear fit over the range $100\leq N\leq2000$ gives $\alpha=1.424$ ($R^2=0.9996$), while restricting the fit to $200\leq N\leq2000$ yields $\alpha=1.447$ (Table~\ref{tab:korcak}). The difference between these estimates indicates some sensitivity to the fitting interval, although both values are consistent with independent estimates of the boundary dimension.

An independent calculation of the uncertainty exponent, based on pairs of randomly selected initial conditions separated by distances $\varepsilon\in[10^{-6.5},10^{-2}]$, gives $\alpha_u=0.571$. For a two-dimensional initial-condition space, this corresponds to a boundary dimension $d_f=2-\alpha_u=1.429$~\cite{mcdonald1985}. These estimates are also consistent with the value $D\approx1.42$ reported by Epureanu and Greenside~\cite{epureanu1998} for the same computational domain.

The resolutions $N<100$ are excluded from the linear fit because the corresponding graphs contain fewer than approximately one hundred vertices, leading to appreciable fluctuations in the local scaling slopes.

The Kor\v{c}\'ak exponent can also be estimated directly from the distribution of component areas on a single grid scale, using Eq.~\eqref{eq:korcak} and the areas $A_i$ associated to each vertex. At the highest resolution, $N=2000$, this procedure gives $2K=1.38$, differing by less than $0.07$ from both estimates of $\alpha$ and from the independently determined boundary dimension. The agreement between these quantities supports the interpretation that, in this example, the proliferation of connected basin components contributes to the observed fractal scaling of the boundary.

\subsection{The relaxed Newton family}

To further assess the vertex-scaling approach, we consider the relaxed Newton family $N_h(z)=z-h\,p(z)/p'(z)$, with $p(z)=z^3+(\lambda-1)z-\lambda$ and $\lambda=1.1-0.1i$. For this family, Saupe computed the box-counting dimension $d_F(h)$ of the Julia set over the parameter range $0.1\leq h\leq1.9$~\cite{saupe1987}. These results provide an independent reference against which the scaling exponents obtained from the basin adjacency graph can be compared.

\begin{figure}[t]
    \centering
    \includegraphics[width=0.9\linewidth]{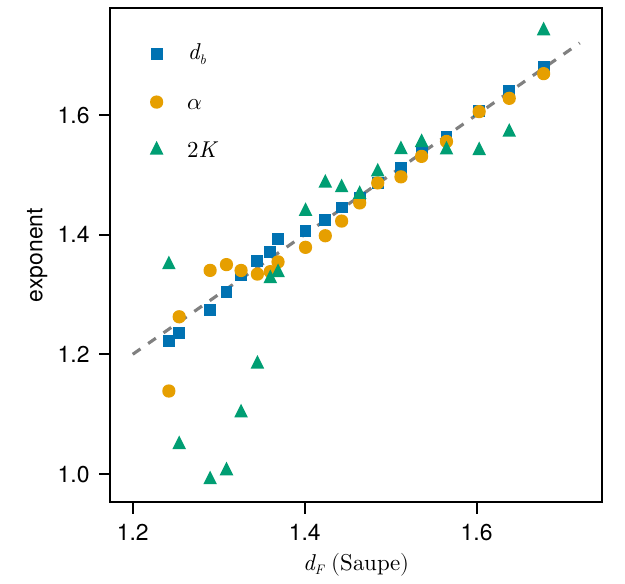}
    \caption{Relaxed Newton family $N_h$,
    $0.1\leq h\leq1.9$: box-counting estimate
    $d_b$ and vertex-scaling exponent $\alpha$,
    obtained using resolutions $N=250$--$2000$,
    together with the single-grid Kor\v{c}\'ak
    estimate $2K$ at $N=2000$, plotted against
    the box-counting dimension $d_F$ reported
    by Saupe~\cite{saupe1987}. The dashed line
    represents the identity.}
    \label{fig:korcak}
\end{figure}

The box-counting estimate $d_b$ agrees with $d_F$ within $0.025$ throughout the parameter range. The vertex-scaling exponent $\alpha$ also shows good agreement, with differences smaller than $0.03$ for $h\geq0.5$ (Fig.~\ref{fig:korcak}). This agreement indicates that, for this family, the proliferation of connected basin components provides a useful measure of the geometrical complexity of the boundary.

The Kor\v{c}\'ak estimate $2K$, obtained directly from the component-area distribution on a single grid, agrees with $\alpha$ within $0.1$ for $h\geq0.7$, corresponding to $d_F\gtrsim1.36$. However, for $h\leq0.6$, the estimate decreases well below $\alpha$, while the vertex-scaling exponent remains close to the independently determined boundary dimension. This discrepancy arises because, in this parameter range, the component-area distribution does not exhibit a well-defined power law over the available range of scales. Consequently, the single-grid Kor\v{c}\'ak estimate becomes unreliable, whereas the resolution-dependent vertex scaling continues to provide a reasonable estimate of the boundary dimension.

\subsection{Flows and maps with thin components}

We now consider three dynamical systems with complex basin structures, including previously reported Wada basins: the magnetic pendulum with three magnets, the quasiperiodically forced map studied by Feudel et al.~\cite{feudel1998}, and the periodically forced two-well Duffing oscillator,
\begin{equation}
    \ddot{x}+\delta\dot{x}-x+x^3=F\sin(\omega t),
\end{equation}
with $\delta=0.15$, $F=0.245$, and $\omega=1$~\cite{aguirre2002,daza2018plos}. For these parameters, the stroboscopic map with period $T=2\pi/\omega$ possesses three coexisting attractors: two period-1 orbits and a period-3 orbit, whose basins exhibit the Wada property~\cite{aguirre2002}. The numerical results for the three systems are summarized in Table~\ref{tab:korcak}.

In the magnetic pendulum and the Feudel map, most resolved basin components consist almost entirely of boundary pixels. The corresponding vertex-scaling exponents, $\alpha=1.85$ and $1.89$, are reasonably close to the box-counting estimates $d_b=1.95$ and $1.93$, respectively. The latter is also consistent with the boundary dimension reported for the Feudel map~\cite{feudel1998}. By contrast, the direct estimates obtained from the component-area distributions on a single grid give $2K=2.42$ and $1.39$, respectively. These discrepancies illustrate the limitations of area-based scaling for thin or highly irregular components, whose detection depends more strongly on their width and spatial arrangement than on their area. In particular, the value $2K=2.42$ cannot be interpreted as a geometrical dimension of a planar boundary. In these examples, the resolution-dependent vertex scaling provides estimates closer to the box-counting dimension of the boundary.

The Duffing oscillator presents a particularly complex basin geometry, with strongly intermingled basin components at the available spatial resolutions, as observed in previous studies~\cite{daza2018plos}. The number of vertices scales with an exponent $\alpha=1.97$, close to the dimension of the two-dimensional initial-condition space. At $N=1000$, approximately $96\%$ of the pixels are classified as boundary pixels, according to the eight-neighbor criterion adopted here. One of the basins forms a large connected component that extends across the sampled domain and is adjacent to almost all the other resolved components. Over the investigated resolution range, the maximum vertex degree exhibits approximately linear growth with $|V|$, with a fitted exponent close to unity.

For this system, the single-grid estimate $2K=2.87$ falls outside the range expected for a geometrical dimension in the plane. This indicates that the assumed relationship between component area and characteristic length scale is not applicable over the available range of resolutions. The result therefore highlights the limitations of interpreting a finite range fit as a geometrical dimension when the basin structure is highly fragmented or insufficiently resolved.

{\setlength{\tabcolsep}{3.5pt}
\begin{table*}[t]
  \centering
  \small
  \caption{Korčák analysis of the BAG. $d_b$ and $\alpha$: exponents of the number of boundary
    pixels and of vertices with the resolution. $2K$: Korčák
    dimension fitted on the finest grid. Wada: range over $N$ of the fraction of boundary
    components adjacent to all other basins; $\bar w$: mean number of other basins adjacent to
    a boundary component, on the finest grid.}
  \label{tab:korcak}
  \begin{tabular}{lcccccc}
    \toprule
    System & $N_A$ & $d_b$ & $\alpha$ & $2K$ & Wada & $\bar w$ \\
    \midrule
    Newton $z^3-1$ & 3 & 1.443 & 1.447 & 1.384 & 1.00--1.00 & 2.00 \\
    Magnetic pendulum & 3 & 1.950 & 1.854 & 2.417 & 0.99--0.99 & 1.99 \\
    Feudel map & 3 & 1.928 & 1.887 & 1.394 & 0.96--0.98 & 1.96 \\
    Duffing oscillator & 3 & 1.986 & 1.966 & 2.871 & 0.98--0.99 & 1.98 \\
    \bottomrule
  \end{tabular}
\end{table*}

}

\subsection{Wada property}

The Wada property requires every boundary point of a basin to belong to the boundaries of all the other basins. Although the BAG does not retain the spatial information necessary to establish this property, it provides a simple measure of the adjacency relations between basin components: the mean basin-neighbor count $\bar w$, obtained by averaging $w(C_i)$ over the boundary components. For basin configurations exhibiting the Wada property, $\bar w$ may approach its maximum value $N_A-1$ as the spatial resolution increases. However, this condition is not sufficient to establish the Wada property, since mutually adjacent components need not share the same boundary.

For Newton's method, $\bar w=2.00$ at $N=2000$, consistent with the independently established Wada property~\cite{daza2018}. Similar values are obtained for the magnetic pendulum ($\bar w=1.99$), the Duffing oscillator ($1.98$), and the Feudel map ($1.96$). Over the investigated resolutions, $\bar w$ fluctuates by less than $0.02$ but shows no systematic trend. Components that do not have neighbors from every basin are generally small and may be affected by finite-resolution effects or by the boundaries of the computational domain.

For the relaxed Newton family, $\bar w$ remains between $1.98$ and $2$ for all values of $h$ studied. These results indicate that the BAG captures characteristic adjacency patterns associated with known Wada basin configurations. Nevertheless, the proposed quantities should be interpreted as graph-based descriptors rather than independent numerical proofs of the Wada property.

\section{Discussion}

The basin adjacency graph (BAG) provides a compact representation of basin structures by retaining the adjacency relations between their connected components. The scaling of the number of graph vertices with spatial resolution characterizes the proliferation of resolved components and, under suitable assumptions, can be related to the Kor\v{c}\'ak exponent of the component-size distribution. This scaling exponent may coincide with the box-counting dimension of the basin boundary when its fractal complexity is associated with a hierarchy of disconnected components. Such a correspondence is illustrated by idealized Sierpi\'nski constructions and is supported numerically by Newton's method and the relaxed Newton family. For sufficiently compact components, the exponent obtained from the area distribution on a single grid agrees reasonably well with the vertex-scaling exponent $\alpha$. However, this agreement breaks down for the thin or highly irregular components encountered in the magnetic pendulum and the Feudel map. In such cases, analyzing the graph over several resolutions provides a more reliable characterization of component proliferation and avoids the dependence on area thresholds required by a direct fit of the size distribution~\cite{imre2012}. The BAG also provides, at little additional computational cost, information about the adjacency of components belonging to different basins. In particular, the mean basin-neighbor count reveals characteristic patterns in systems with known Wada basins. Nevertheless, this quantity does not constitute an independent test of the Wada property, which requires every boundary point to be shared by all the basins.

The method has several limitations. First, fractal boundaries separating a finite number of connected regions, such as quasicircle Julia sets, may produce graphs with a fixed number of vertices, independently of the geometrical complexity of the boundary. In these cases, the BAG does not capture the roughness of an individual boundary, and complementary methods, such as the uncertainty exponent or basin entropy~\cite{mcdonald1985,daza2016,gusso2021}, remain necessary. Second, estimating the vertex-scaling exponent requires basin diagrams at several spatial resolutions, which may be difficult when only a single experimental image is available. Although coarser representations can sometimes be obtained from such an image, the accessible scaling range is necessarily limited by its original resolution. Third, the graph retains adjacency information but not the spatial arrangement of the interfaces between components. Consequently, local properties of the basin boundaries, including the pointwise Wada condition, cannot be determined from component adjacency alone. Moreover, the resulting graph depends on the connectivity convention adopted for the sampled grid, particularly when small components or diagonal contacts are present.

The BAG belongs to the broader family of graph-based representations of dynamical systems~\cite{donner2010,zou2019,deleo2021}, but differs from conventional fractal-network approaches. The latter typically characterize scaling by covering a graph with boxes defined in terms of graph distance~\cite{song2007,bunimovich2024}, possibly with variable box sizes~\cite{lepek2025}, or by studying relationships between the degree of highly connected vertices and the mass of graph boxes~\cite{fronczak2024}. In contrast, the scaling properties of the BAG originate from the spatial organization and proliferation of basin components in the underlying two-dimensional domain. Its characteristic scale is therefore the spatial resolution at which these components can be distinguished, rather than a distance defined on the graph itself.

The degree distribution provides additional information about the adjacency structure. Small components often have relatively few neighbors, whereas large connected components may be adjacent to many smaller ones, producing high-degree vertices. In the systems considered, the mean degree remains between approximately 4 and 6 for Newton's method, the magnetic pendulum, and the Feudel map, despite the presence of vertices with much larger degrees. Although the degree distribution may exhibit a broad or approximately power-law tail, its scaling cannot be inferred directly from the Kor\v{c}\'ak law. The number of neighboring components depends not only on their sizes but also on their spatial arrangement and mutual contacts. Thus, the degree distribution contains information about the organization of the basin components that is not determined by the component-area distribution alone. Exploring this relationship more systematically could provide an interesting direction for future work.

Finally, the Kor\v{c}\'ak exponent has also been used to characterize the patchiness of landscapes and habitat fragmentation~\cite{hastings1982,imre2012}. The BAG could therefore offer a complementary description of categorical spatial maps, providing information about the connectivity and adjacency of different regions beyond their size distributions. This suggests potential applications of the approach outside nonlinear dynamics.

\section*{Numerical Methods}

The basins of attraction are computed using the Julia package \texttt{Attractors.jl}~\cite{datseris2023}. For each system, the attractors are identified before constructing the basin diagrams. They are obtained either through the recurrence-based algorithm, as in the Duffing oscillator and the Feudel map, or from the known attracting solutions, as in Newton's method and the magnetic pendulum. Each initial condition on a regular $N\times N$ grid is then assigned to the attractor approached by its trajectory. The same attractor labels are used at different spatial resolutions to ensure consistency in the comparison of basin diagrams.

Continuous-time systems are integrated using the adaptive Vern9 scheme, with numerical tolerances of $10^{-8}$. For the periodically forced Duffing oscillator, trajectories are sampled stroboscopically with period $T=2\pi/\omega$.

The parameters and computational domains are as follows. For Newton's method, the initial conditions are sampled on $[-1,1]^2$, while the relaxed Newton family is studied on $[-2,2]^2$. For the Duffing oscillator, we use $\delta=0.15$, $F=0.245$, and $\omega=1$, with initial conditions $(x,\dot{x})\in[-2,2]^2$. The magnetic pendulum is described by
\begin{equation}
\ddot{\mathbf r}
=
-\omega_0^2\mathbf r
-b\dot{\mathbf r}
+\sum_{i=1}^{3}
\frac{\gamma_m(\mathbf r_i-\mathbf r)}
{\left(|\mathbf r_i-\mathbf r|^2+h^2\right)^{3/2}},
\end{equation}
where the three magnets are located at positions $\mathbf r_i$ on the unit circle. We take $\gamma_m=1$, $h=0.3$, $\omega_0=0.5$, and $b=0.05$. The pendulum is released from rest, with initial positions $\mathbf r\in[-2,2]^2$. For the quasiperiodically forced Feudel map, we use $r=3.833$ and $a=0.0015$, with initial conditions $(u,x)\in[0,1]^2$.

For Newton's method, the vertex-scaling fit reported in Table~\ref{tab:korcak} uses $200\leq N\leq2000$. Figure~\ref{fig:newton}(b) also includes coarser grids, including resolutions below $N=100$, and reports a fit over $100\leq N\leq2000$. Points with $N<100$ are excluded from that fit. For the magnetic pendulum and the Feudel map, we use $N=100$--$500$, while for the Duffing oscillator we consider $N=100$--$1000$. The relaxed Newton family is analyzed over $N=250$--$2000$, using an approximately geometric sequence of resolutions with successive values increasing by a factor of $\sqrt{2}$.

The connected components of each basin are identified using eight-neighbor pixel connectivity, implemented through a union--find algorithm. The resulting basin adjacency graphs are constructed and analyzed using \texttt{Graphs.jl}. The vertex-scaling exponent $\alpha$ is estimated from a least-squares fit of $\log|V|$ against $\log N$. As a reference for the boundary dimension, we use a box-counting estimate $d_b$, obtained from the scaling with $N$ of the number of boundary pixels, defined as pixels having at least one eight-neighbor belonging to a different basin. This corresponds to box counting with box size $\varepsilon=1/N$. The component-area distribution is also analyzed on the finest grid to estimate the Kor\v{c}\'ak exponent $K$, following the fitting procedure of Imre et al.~\cite{imre2012}. The fit uses component areas between 10 pixels and $10^{-3}$ of the total image area.

For comparison, we also consider independent estimates of the boundary dimension obtained from the uncertainty exponent~\cite{mcdonald1985}, as well as the merging-method criterion for identifying Wada basin boundaries~\cite{daza2018}. The adjacency statistics extracted from the BAG are interpreted separately from these independent tests, since component adjacency alone does not establish the Wada property.

The code that reproduces all figures and tables is available at \url{https://github.com/awage/AnalysisBasinsNets}.

\par\vspace{1\baselineskip}

\begin{acknowledgments}
This work was supported by the Spanish State Research Agency (AEI) and the European Regional Development Fund (ERDF, European Union) under Project No.~PID2023-148160NB-I00 (MCIN/AEI/10.13039/501100011033).
\end{acknowledgments}


\begin{thebibliography}{99}
\providecommand{\natexlab}[1]{#1}
\providecommand{\doi}[1]{\href{https://doi.org/#1}{doi:\,#1}}

\bibitem[Daza et~al.(2024)Daza, Wagemakers, and Sanju{\'a}n]{daza2024pt}
{\'A}lvar Daza, Alexandre Wagemakers, and Miguel A.~F. Sanju{\'a}n.
\newblock Multistability and unpredictability.
\newblock \emph{Phys. Today}, 77(11):44--50, 2024.
\newblock \doi{10.1063/pt.kcxv.poxf}.

\bibitem[Grebogi et~al.(1983)Grebogi, McDonald, Ott, and Yorke]{grebogi1983}
Celso Grebogi, Steven~W. McDonald, Edward Ott, and James~A. Yorke.
\newblock Final state sensitivity: An obstruction to predictability.
\newblock \emph{Phys. Lett. A}, 99(9):415--418, 1983.

\bibitem[McDonald et~al.(1985)McDonald, Grebogi, Ott, and Yorke]{mcdonald1985}
Steven~W. McDonald, Celso Grebogi, Edward Ott, and James~A. Yorke.
\newblock Fractal basin boundaries.
\newblock \emph{Physica D}, 17(2):125--153, 1985.
\newblock \doi{10.1016/0167-2789(85)90001-6}.

\bibitem[Aguirre et~al.(2009)Aguirre, Viana, and Sanju{\'a}n]{aguirre2009}
Jacobo Aguirre, Ricardo L. Viana, and Miguel A.~F. Sanju{\'a}n.
\newblock Fractal structures in nonlinear dynamics.
\newblock \emph{Rev. Mod. Phys.}, 81:333--386, 2009.
\newblock \doi{10.1103/RevModPhys.81.333}.

\bibitem[Daza et~al.(2016)Daza, Wagemakers, Georgeot, Gu{\'e}ry-Odelin, and
  Sanju{\'a}n]{daza2016}
Alvar Daza, Alexandre Wagemakers, Bertrand Georgeot, David Gu{\'e}ry-Odelin, and Miguel A.~F. Sanju{\'a}n.
\newblock Basin entropy: a new tool to analyze uncertainty in dynamical systems.
\newblock \emph{Sci. Rep.}, 6:31416, 2016.
\newblock \doi{10.1038/srep31416}.

\bibitem[Daza et~al.(2022)Daza, Wagemakers, Georgeot, Gu{\'e}ry-Odelin, and
  Sanju{\'a}n]{daza2022}
Alvar Daza, Alexandre Wagemakers, Bertrand Georgeot, David Gu{\'e}ry-Odelin, and Miguel A.~F. Sanju{\'a}n.
\newblock Classifying basins of attraction using the basin entropy.
\newblock \emph{Chaos Solitons Fractals}, 159:112112, 2022.
\newblock \doi{10.1016/j.chaos.2022.112112}.

\bibitem[Gusso and de~Mello(2021)]{gusso2021}
Andr{\'e} Gusso and Lucas~E. de~Mello.
\newblock Fractal dimension of basin boundaries calculated using the basin entropy.
\newblock \emph{Chaos Solitons Fractals}, 153:111532, 2021.
\newblock \doi{10.1016/j.chaos.2021.111532}.

\bibitem[Tr{\'e}meau and Colantoni(2000)]{tremeau2000}
Alain Tr{\'e}meau and Philippe Colantoni.
\newblock Regions adjacency graph applied to color image segmentation.
\newblock \emph{IEEE Trans. Image Process.}, 9(4):735--744, 2000.
\newblock \doi{10.1109/83.841950}.

\bibitem[Donner et~al.(2010)Donner, Zou, Donges, Marwan, and
  Kurths]{donner2010}
Reik~V. Donner, Yong Zou, Jonathan~F. Donges, Norbert Marwan, and J{\"u}rgen Kurths.
\newblock Recurrence networks---a novel paradigm for nonlinear time series analysis.
\newblock \emph{New J. Phys.}, 12:033025, 2010.
\newblock \doi{10.1088/1367-2630/12/3/033025}.

\bibitem[Zou et~al.(2019)Zou, Donner, Marwan, Donges, and Kurths]{zou2019}
Yong Zou, Reik~V. Donner, Norbert Marwan, Jonathan~F. Donges, and J{\"u}rgen Kurths.
\newblock Complex network approaches to nonlinear time series analysis.
\newblock \emph{Phys. Rep.}, 787:1--97, 2019.
\newblock \doi{10.1016/j.physrep.2018.10.005}.

\bibitem[De~Leo and Yorke(2021)]{deleo2021}
Roberto De~Leo and James~A. Yorke.
\newblock The graph of the logistic map is a tower.
\newblock \emph{Discrete Contin. Dyn. Syst.}, 41(11):5243--5269, 2021.
\newblock \doi{10.3934/dcds.2021075}.

\bibitem[Mandelbrot(1982)]{mandelbrot1982}
Benoit~B. Mandelbrot.
\newblock \emph{The Fractal Geometry of Nature}.
\newblock Freeman, New York, 1982.

\bibitem[Hastings et~al.(1982)Hastings, Pekelney, Monticciolo, Vun~Kannon, and
  Del~Monte]{hastings1982}
Harold~M. Hastings, Richard Pekelney, Richard Monticciolo, David Vun~Kannon, and Diane Del~Monte.
\newblock Time scales, persistence and patchiness.
\newblock \emph{Biosystems}, 15:281--289, 1982.

\bibitem[Imre et~al.(2012)Imre, Novotn{\'y}, and Rocchini]{imre2012}
Attila~R. Imre, Josef Novotn{\'y}, and Duccio Rocchini.
\newblock The {K}orcak-exponent: A non-fractal descriptor for landscape patchiness.
\newblock \emph{Ecol. Complex.}, 12:70--74, 2012.
\newblock \doi{10.1016/j.ecocom.2012.10.001}.

\bibitem[Kennedy and Yorke(1991)]{kennedy1991}
Judy Kennedy and James A. Yorke.
\newblock Basins of {W}ada.
\newblock \emph{Physica D}, 51:213--225, 1991.
\newblock \doi{10.1016/0167-2789(91)90234-Z}.

\bibitem[Nusse and Yorke(1996)]{nusse1996}
Helena~E. Nusse and James~A. Yorke.
\newblock Wada basin boundaries and basin cells.
\newblock \emph{Physica D}, 90(3):242--261, 1996.
\newblock \doi{10.1016/0167-2789(95)00249-9}.

\bibitem[Daza et~al.(2015)Daza, Wagemakers, Sanju{\'a}n, and Yorke]{daza2015}
Alvar Daza, Alexandre Wagemakers, Miguel A.~F. Sanju{\'a}n, and James~A. Yorke.
\newblock Testing for basins of {W}ada.
\newblock \emph{Sci. Rep.}, 5:16579, 2015.
\newblock \doi{10.1038/srep16579}.

\bibitem[Wagemakers et~al.(2021)Wagemakers, Daza, and
  Sanju{\'a}n]{wagemakers2021}
Alexandre Wagemakers, Alvar Daza, and Miguel A.~F. Sanju{\'a}n.
\newblock How to detect {W}ada basins.
\newblock \emph{Discrete Contin. Dyn. Syst. Ser. B}, 26(1):717--739, 2021.
\newblock \doi{10.3934/dcdsb.2020330}.

\bibitem[Yoneyama(1917)]{yoneyama1917}
Kuniz{\^o} Yoneyama.
\newblock Theory of continuous set of points (not finished).
\newblock \emph{Tohoku Math. J.}, 12:43--158, 1917.

\bibitem[Rosenfeld(1979)]{rosenfeld1979}
Azriel Rosenfeld.
\newblock Digital topology.
\newblock \emph{Am. Math. Mon.}, 86(8):621--630, 1979.
\newblock \doi{10.1080/00029890.1979.11994873}.

\bibitem[Farmer et~al.(1983)Farmer, Ott, and Yorke]{farmer1983}
J.~Doyne Farmer, Edward Ott, and James~A. Yorke.
\newblock The dimension of chaotic attractors.
\newblock \emph{Physica D}, 7:153--180, 1983.

\bibitem[Theiler(1990)]{theiler1990}
James Theiler.
\newblock Estimating fractal dimension.
\newblock \emph{J. Opt. Soc. Am. A}, 7(6):1055--1073, 1990.

\bibitem[Epureanu and Greenside(1998)]{epureanu1998}
Bogdan~I. Epureanu and Henry~S. Greenside.
\newblock Fractal basins of attraction associated with a damped {N}ewton's method.
\newblock \emph{SIAM Rev.}, 40:102--109, 1998.

\bibitem[Saupe(1987)]{saupe1987}
Dietmar Saupe.
\newblock Efficient computation of {J}ulia sets and their fractal dimension.
\newblock \emph{Physica D}, 28:358--370, 1987.

\bibitem[Feudel et~al.(1998)Feudel, Witt, Lai, and Grebogi]{feudel1998}
Ulrike Feudel, Annette Witt, Ying-Cheng Lai, and Celso Grebogi.
\newblock Basin bifurcation in quasiperiodically forced systems.
\newblock \emph{Phys. Rev. E}, 58(3):3060--3066, 1998.
\newblock \doi{10.1103/PhysRevE.58.3060}.

\bibitem[Aguirre and Sanju{\'a}n(2002)]{aguirre2002}
Jacobo Aguirre and Miguel A.~F. Sanju{\'a}n.
\newblock Unpredictable behavior in the {D}uffing oscillator: {W}ada basins.
\newblock \emph{Physica D}, 171(1--2):41--51, 2002.
\newblock \doi{10.1016/S0167-2789(02)00565-1}.

\bibitem[Daza et~al.(2018{\natexlab{a}})Daza, Wagemakers, and
  Sanju{\'a}n]{daza2018plos}
Alvar Daza, Alexandre Wagemakers, and Miguel A.~F. Sanju{\'a}n.
\newblock From local uncertainty to global predictions: Making predictions on fractal basins.
\newblock \emph{PLoS ONE}, 13(4):e0194926, 2018a.
\newblock \doi{10.1371/journal.pone.0194926}.

\bibitem[Daza et~al.(2018{\natexlab{b}})Daza, Wagemakers, and
  Sanju{\'a}n]{daza2018}
Alvar Daza, Alexandre Wagemakers, and Miguel A.~F. Sanju{\'a}n.
\newblock Ascertaining when a basin is {W}ada: the merging method.
\newblock \emph{Sci. Rep.}, 8:9954, 2018b.
\newblock \doi{10.1038/s41598-018-28119-0}.

\bibitem[Song et~al.(2007)Song, Gallos, Havlin, and Makse]{song2007}
Chaoming Song, Lazaros~K. Gallos, Shlomo Havlin, and Hern{\'a}n~A. Makse.
\newblock How to calculate the fractal dimension of a complex network: the box covering algorithm.
\newblock \emph{J. Stat. Mech.: Theory Exp.}, 2007(03):P03006, 2007.
\newblock \doi{10.1088/1742-5468/2007/03/P03006}.

\bibitem[Bunimovich and Skums(2024)]{bunimovich2024}
Leonid Bunimovich and Pavel Skums.
\newblock Fractal networks: Topology, dimension, and complexity.
\newblock \emph{Chaos}, 34(4):042101, 2024.
\newblock \doi{10.1063/5.0200632}.

\bibitem[{\L}epek et~al.(2025){\L}epek, Makulski, Fronczak, and
  Fronczak]{lepek2025}
Micha{\l} {\L}epek, Kordian Makulski, Agata Fronczak, and Piotr Fronczak.
\newblock Beyond traditional box-covering: Determining the fractal dimension of complex networks using a fixed number of boxes of flexible diameter.
\newblock \emph{Chaos Solitons Fractals}, 199:116908, 2025.
\newblock \doi{10.1016/j.chaos.2025.116908}.

\bibitem[Fronczak et~al.(2024)Fronczak, Fronczak, Samsel, Makulski, {\L}epek,
  and Mrowinski]{fronczak2024}
Agata Fronczak, Piotr Fronczak, Mateusz~J. Samsel, Kordian Makulski, Micha{\l} {\L}epek, and Maciej~J. Mrowinski.
\newblock Scaling theory of fractal complex networks.
\newblock \emph{Sci. Rep.}, 14:9079, 2024.

\bibitem[Datseris et~al.(2023)Datseris, Rossi, and Wagemakers]{datseris2023}
George Datseris, Kalel~Luiz Rossi, and Alexandre Wagemakers.
\newblock Framework for global stability analysis of dynamical systems.
\newblock \emph{Chaos}, 33(7):073151, 2023.
\newblock \doi{10.1063/5.0159675}.

\bibitem[Datseris and Parlitz(2022)]{datseris2022}
George Datseris and Ulrich Parlitz.
\newblock \emph{Nonlinear Dynamics: A Concise Introduction Interlaced with Code}.
\newblock Springer, Cham, 2022.
\newblock \doi{10.1007/978-3-030-91032-7}.

\end{thebibliography}

\end{document}